\documentclass[transaction]{IEEEtran}
\usepackage{graphicx}
\usepackage{enumitem}
\usepackage{subcaption}
\usepackage{amsmath,slashbox}
\usepackage{multirow}
\usepackage{mathtools}
\usepackage{balance}
\usepackage{booktabs}
\usepackage{breqn}
\usepackage[noadjust]{cite}
\usepackage{color}

\usepackage[]{algorithm2e}

\begin{document}

\title{Learning from Natural Noise to Denoise Micro-Doppler Spectrogram}

\author{
    \IEEEauthorblockN{Chong Tang\IEEEauthorrefmark{1}, Wenda Li\IEEEauthorrefmark{1}, Shelly Vishwakarma\IEEEauthorrefmark{1}, Karl Woodbridge\IEEEauthorrefmark{3}, Simon Julier\IEEEauthorrefmark{2}, Kevin Chetty\IEEEauthorrefmark{1}}
    \IEEEauthorblockA{\\\IEEEauthorrefmark{1}Department of Security and Crime Science, University College London, UK}\\
    \IEEEauthorblockA{\IEEEauthorrefmark{2}Department of Computer Science, University College London, UK}\\
    \IEEEauthorblockA{\IEEEauthorrefmark{3}Department of Electronic\&Electrical Engineering, University College London, UK}
    \\{\IEEEauthorrefmark{1}{{chong.tang.18, wenda.li, s.vishwakarma, k.chetty}}@ucl.ac.uk,\IEEEauthorrefmark{2}{s.julier}@ucl.ac.uk, \IEEEauthorrefmark{3}{k.woodbridge}@ucl.ac.uk}
}

\maketitle

\begin{abstract}
Micro-Doppler analysis has become increasingly popular in recent years owning to the ability of the technique to enhance classification strategies. Applications include recognising everyday human activities, distinguishing drone from birds, and identifying different types of vehicles. However, noisy time-frequency spectrograms can significantly affect the performance of the classifier and must be tackled using appropriate denoising algorithms. In recent years, deep learning algorithms have spawned many deep neural network-based denoising algorithms. For these methods, noise modelling is the most important part and is used to assist in training. In this paper, we decompose the problem and propose a novel denoising scheme: first, a Generative Adversarial Network (GAN) is used to learn the noise distribution and correlation from the real-world environment; then, a simulator is used to generate clean Micro-Doppler spectrograms; finally, the generated noise and clean simulation data are combined as the training data to train a Convolutional Neural Network (CNN) denoiser. In experiments, we qualitatively and quantitatively analyzed this procedure on both simulation and measurement data. Besides, the idea of learning from natural noise can be applied well to other existing frameworks and demonstrate greater performance than other noise models.

\end{abstract}


\IEEEpeerreviewmaketitle

\section{Introduction}
\label{Introduction}
The Micro-Doppler effect~\cite{chen2006micro} refers to that fact that, in radar or sonar systems, the returned information contains not only the bulk motion but also micro-motions of any structure on the target such as arm swinging and breathing. In many research fields such as human activity recognition~\cite{kim2009human,li2017passive}, hand gesture recognition~\cite{kim2016hand}, and occupancy detection~\cite{tang2020occupancy}, the micro-Doppler spectrogram (MDS) is used to express micro-Doppler information in time-frequency domain. In recent years, because of the low-cost, high-sensitivity and non-cooperative characteristics, MDS-based sensing technologies have  attracted significant  attention. On the other hand, the performance of the technology is positively related to the quality of the spectrogram. Chen et al~\cite{chen2010target} demonstrate that, as the signal-to-noise ratio (SNR) increases, the target recognition rate becomes better. Therefore, to obtain high-quality MDS, we need to mitigate the effect caused by interference and noise. Algorithms based on Principle Component Analysis~\cite{du2013robust} and CLEAN technique~\cite{smith2008naive} can be adopted. Although they can remove significant interference like direct signal interference and noise, there is some residual noise, which still degrades the performance of results. Therefore, an effective denoising method is crucial for the future development of wireless sensing applications.

The success of deep neural networks (DNNs) in the field of image processing has inspired emergence of various DNN-based MDS denoising methods. Researchers in \cite{rock2020deep}\cite{rock2019complex} propose a CNN-based technique for denoising FMCW radar MDS. Two trained CNN models are applied at different steps in the radar signal processing, and the denoising performance is tested on both simulated and measured data. Another study, \cite{de2020deep} utilized the feature extraction ability of CNNs to design Deep Convolutional Autoencoders (CAE). Experimental results indicated that the CAE method outperforms traditional Constant False Alarm Rate method at any noise level. In recent years, GANs\cite{goodfellow2014generative} have also become popular in the field of the radar signal denoising, and has been applied in many scenarios. Huang et al~\cite{huang2018micro} leverages the noisy spectrogram and corresponding clean ground truth to train the GAN which is able to output the denoised spectrogram. Finally, the trained generator is able to directly generate the denoised MDS according to the noisy input. In contrast, \cite{he2020deep} is an indirect method for noise reduction. The work first uses a fully convolutional neural networks to localize the noisy region, then the GAN which is trained with clean MDS removes those regions and restores them according to the distribution of clean data. 

These approaches have shown the huge potential of learning-based algorithms for radar signal denoising. Moreover, due to the excellent feature extraction and information generation abilities, they sometimes outperform traditional methods. Nevertheless, no matter what kind of DNN-based methods, they require sufficient noisy data and the corresponding clean data to train the networks. However, capturing clean data is quite challenging for wireless sensing experiments. Therefore, most existing studies use the simulated data and artificially added noise to create a training set. However, the assumption of this artificial noise usually is Additive White Gaussian Noise (AWGN), which is pixel-independent and rarely matches the spatially-correlated real-world noise. To extensively apply learning-based algorithms in more complex real-world scenarios, the synthetic noise should close to the real noise distribution as much as possible. So we need to replace AWGN with a more sophisticated noise.

In the field of image processing, various methods have been proposed to deal with similar issues. To better simulate the real noise, the Pixel-shuffle Down-sampling strategy~\cite{zhou2020awgn} can be used to decompose the spatially-correlated real noise into spatially-variant and pixel-independent noise terms so that AWGN can locally approximate noise distribution. However, this technique might not be suitable for denoising radar signals. Noise in MDS comes from various sources such as multi-path propagation clutter, other RF sources, etc. Therefore, the noise in MDS can have a relatively long-term temporal correlation and relatively large-scale spatial correlation. Therefore, this type of noise should be synthesized according to the global distribution rather than approximated using AWGN from locally to globally. In this case,~\cite{chen2018image} provides us with a starting point. They utilized a GAN to learn complex distribution from natural noise to then generate more realistic noise. Then the synthetic noise is added on clean images to create a paired-image dataset for training CNN-based denoising networks. Although there are some limitations as mentioned in \cite{tran2020gan}, for MDS denoising, we can adopt GAN to create a close-to-natural noise distribution. Furthermore, based on this distribution, we can generate an unlimited amount of training pairs.

In this paper, we propose a GAN-based noise modelling method to denoise MDS. We first use GAN to model the noise distribution from experimentally measured spectrograms. This is used to generate a close-to-natural noise signal which can be added to clean simulated spectrograms. After that, we obtain a sufficient training set to train a denoising convolutional neural network (DnCNN)\cite{zhang2017beyond} denoiser. In experiments, we extensively validate our method on simulated data and different measurement data sets. By comparing the denoised spectrograms with algorithms that use AWGN, we can see the superiority of our approach qualitatively and quantitatively. Meanwhile, compared with existing GAN-based methods, our method only need to learn the noise distribution. In fact, we simplify the learning process and obtain a reliable model faster. The major contributions of this work are:
\begin{enumerate}
    \item We apply a GAN-based method to learn the real noise distribution and generate improved training pairs. In this way, the CNN-based denoiser can be trained with more realistic data to improve the denoising effect.
    \item Aside from the denoising, our approach provides a strategy for improving simulated spectrograms. The synthetic noise has environmental factors which can make simulated data more realistic after adding the noise. Examples are shown in Section~\ref{Algorithm Framework}.
\end{enumerate}
Our paper is organized as follows. Section \ref{Algorithm Framework} describes how GAN learns from the natural noise to generate a noise patch, and how we create a paired training data by adding the noise patch on the simulated data. After that, it also presents the details of the applied DnCNN denoiser. Section \ref{Experimental Setting} explains how we design our experiments to demonstrate our approach. Finally, Sections \ref{Results Discussion} and \ref{Conclusion} present the experimental results and discuss the denoising performance of the proposed framework.

\section{Denoising Framework}
\label{Algorithm Framework}
\begin{figure}
\centering
\includegraphics[width=\linewidth]{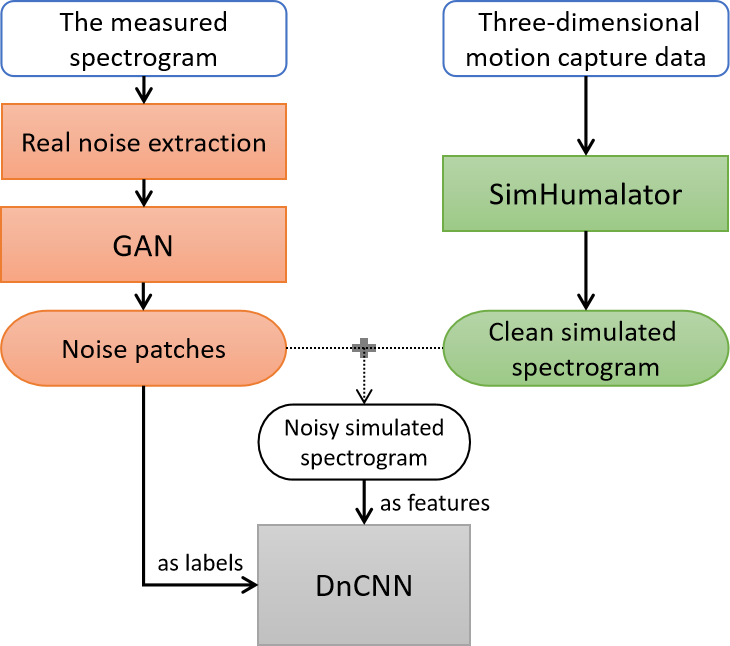}
\caption{Denoising Algorithm Framework}
\label{fig:framework}
\end{figure}

Fig. \ref{fig:framework} presents the overview of our approach. In this section, we introduce the framework details of the proposed method. Our approach consists of two parts --- noise patch generation and denoiser training. Noise patch generation is carried out using GANs. We explain the training methodology of the network, and then describe how we create a training pair according to a noise patch and a simulated data. As mentioned before, our approach trains GAN to produce the noise rather than the denoised spectrogram. For denoiser training part, we utilized DnCNN structure to train a denoiser, where the training pairs are obtained from the last step. Specifically for this study, the experimental data collection and simulated data generation are based on Passive WiFi Radar (PWR) system~\cite{li2020passive} and SimHumalator~\cite{vishwakarmasimhumalator}, respectively. 
\subsection{Noise Patch Generation}
\subsubsection{Real Noise Extraction}
\begin{figure}
\centering
\includegraphics[width=\linewidth]{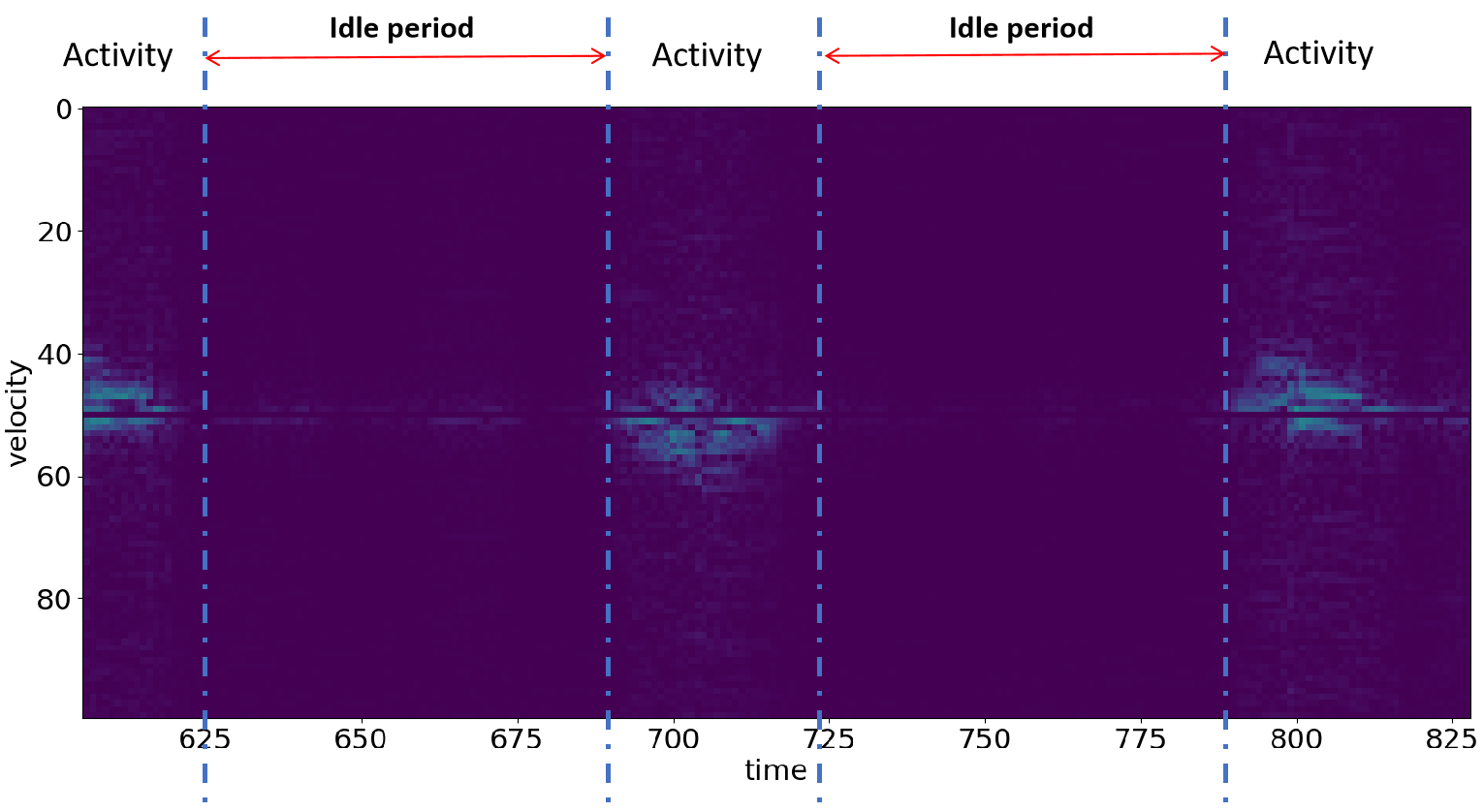}
\caption{An example period of the activity recognition dataset}
\label{fig:fig2}
\end{figure}

Extracting real noise is the primary step for training the GAN. For a noisy spectrogram $\boldsymbol{S}$ which includes target motion information $\boldsymbol{M}$ and noise $\boldsymbol{N}$, it is difficult to separate $\boldsymbol{N}$ from $\boldsymbol{N}+\boldsymbol{M}$. In this case, we can use regions only having $N$ from the recording. Alternatively, we can record some empty background at the beginning of the experiment prior to the target entering to the scene. These "empty" spectrograms contain various environmental interference and noise for GAN to learn. As shown in Fig. \ref{fig:fig2}, there is a short idle period between every two activity recordings. We use an n-by-m window to extract real noise patches from idle periods. We denote a real noise patch as $R$.
\subsubsection{Learning with GAN}
\begin{figure*}
\centering
\includegraphics[scale=0.4]{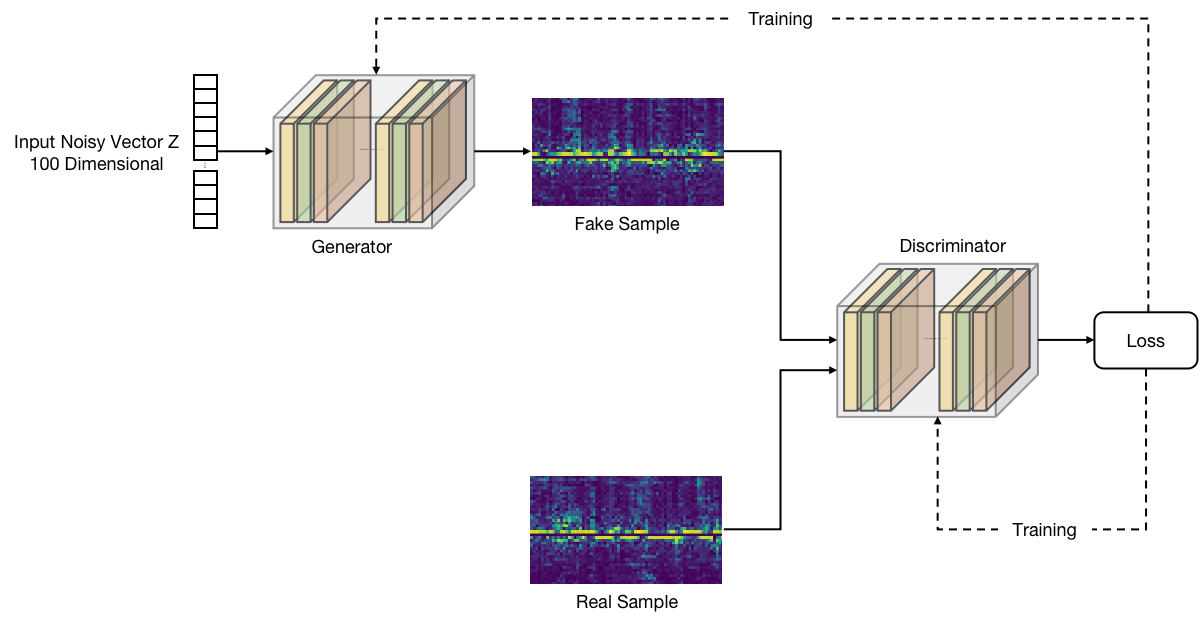}
\caption{The network overview of WGAN-GP}
\label{fig:fig3}
\end{figure*}

The details of the GAN network are shown in Fig. \ref{fig:fig3}. Many papers  have shown that GANs are able to learn complex distributions, which is due to its unique structure and training mechanism~\cite{yu2017seqgan,chen2016infogan,odena2017conditional}. A basic GAN structure includes a generative model G (as known as generator) and a discriminative model D (as known as discriminator). The two models are trained simultaneously but also compete with each other. The generator aims to fool the discriminator by producing close-to-real data. By contrast, the discriminator learns distribution from the real data to determine whether the produced sample from the generator is real enough. Through this minimax two-player game, the generator is able to recover the real data distribution and produce samples according to it. We can express the cost function of the GAN with:
\begin{dmath}
\label{basic_gan_loss}
\mathop{minmax}_{\text{G    D}}L=E_{x\sim p_{r}(x)}[logD(x)]\\+E_{x\sim p_{g}(x)}[log(1-D(x))] 
\end{dmath}
where $p_r$ represents data distribution over real samples and $p_g$ is data distribution learned by the generator. Studies such as \cite{huang2018micro} have already used this process in the MDS denoising. However, optimizing this cost function has some weaknesses during the training. For example, it is hard to train a GAN to achieve the Nash equilibrium~\cite{salimans2016improved}. Another frequently occurring problem is that when the discriminator is trained perfectly, the cost function $\boldsymbol{L}$ becomes zero so that there is no gradient to update the loss, known as a vanishing gradient. Therefore, our network applies WGAN-GP~\cite{gulrajani2017improved} which is an improved version of the previous GAN to effectively avoid these problems. The cost function of the WGAN-GP is:
\begin{dmath}
\label{wgangp_loss}
L=E_{\tilde{x}\sim p_{g}}[D(x)]-E_{x\sim p_{r}}[D(x)]+\lambda E_{\hat{x}\sim p_{\hat{x}}}\left[(||\nabla_{\hat{x}}D(\hat{x})||_{2}-1)^{2}\right]
\end{dmath}
where, in our network, $p_r$ represents data distribution over real noise and $p_g$ is noise distribution learned by the generator and $p_{\hat{x}}$ is uniformly sampled along straight lines between pairs of points sampled from $p_r$ and $p_g$. The WGAN-GP uses the Wasserstein distance as a better metric of distribution similarity, and improves the WGAN~\cite{salimans2016improved} performance by replacing the weight clipping with the gradient penalty term so that we can obtain more stable training process and higher quality synthetic noise. Meanwhile, we use CNNs for the generator and discriminator as suggested in~\cite{radford2015unsupervised} for better learn features from the image-like MDS. 

Eventually, we can use the trained generator to produce the noise patch, $P$.

\subsection{Training Micro-Doppler Spectrogram Denoiser}
\subsubsection{Training Pairs Generation}
According to the framework in Fig. \ref{fig:framework}, a training pair for DnCNN is [Noisy simulated spectrogram, Noise patch]. The noisy simulated spectrogram is obtained after we get generated noise patches from GAN, and we add them to clean simulated data to construct this form of training pairs.

We denote a clean simulated spectrograms as $S$, where $S$ has the same size as $R$ and $P$. A noisy simulated spectrogram, $S_N$, can be obtained by $S+P$, and the corresponding training pair is $[S_N, P]$. Furthermore,by adding $n$ different $P$ to the same $S$, we will get $n$ different $S_N$. Theoretically, we can create an unlimited number of training pairs with this augmentation method. 
\subsubsection{Training with DnCNN}
DnCNN plays the role of completing the denoising process. On the other hand, many other CNN architectures like UNet\cite{ronneberger2015u} can also be used. But for our focus, the characteristics of DnCNN could be more helpful. It adopts residual learning to effectively separate noise from noisy input. Meanwhile, residual learning integrated with the batch normalization can speed up the training process as well as boost the denoising performance. The objective function of DnCNN in our framework is:
\begin{dmath}
\label{wgangp_loss}
L(\Theta)=\frac{1}{2N}\sum_{i=1}^{N}||R(S_{N_i};\Theta)-P_i||_{F}^{2}
\end{dmath}
where $\Theta$ is the vector of trainable parameters in DnCNN, $N$ is the size of training data, $S_{N_i}$ and $P_i$ are from the $i^{th}$ training pair and $R(S_{N_i};\Theta)$ is estimated output of the network. So, unlike other CNN-based architectures, DnCNN aims to estimate noise patch from the noisy input rather than directly output a denoised spectrogram, which greatly reduces the complexity of learning. 

In experiments, $S_{N_i}$ is a noisy simulated spectrogram and $P_i$ is the corresponding noise patch for training the DnCNN. The network structure has been shown in Fig. \ref{fig:fig4}. Moreover, ReLU~\cite{he2015delving} is also applied to improve the training of the network.

\section{Experimental Setting}
\label{Experimental Setting}
In this section, we introduce some fundamental aspects of our experiments, including what metrics have been used to evaluate the denoising performance, data collection, parameter settings for the GAN and the DnCNN and different noise models for comparison.
\subsection{Structural Similarity Index Measure}
The structural similarity index measure (SSIM)~\cite{wang2004image} is used for measuring the structural similarity between two images. Unlike mean square error or peak signal-to-noise ratio estimates absolute errors, the SSIM is a perception-based model that considers image degradation as perceived changes in structural information, which is more consistent with the assessment of the human visual system, and the MDS of different activities has identifiable visual pattern. Therefore, it is suitable to use SSIM to evaluate MDS denoising performance. 

\begin{table}[]
\resizebox{\linewidth}{!}{%
\begin{tabular}{|c|c|c|}
\hline
Layer            & Parameters                                                & Output Data Size \\ \hline
Input            & 100                                                       &                  \\ \hline
Dense            & 22400, BatchNormalization, LeakyReLU                      & 22400            \\ \hline
Reshape          & (25, 7, 128)                                              & (25, 7, 128)     \\ \hline
TransposedConv2D & 128, 5x5, Strides 1, Pad 0, BatchNormalization, LeakyReLU & (25, 7, 128)     \\ \hline
TransposedConv2D & 64, 3x3, Strides 1, Pad 0, BatchNormalization, LeakyReLU  & (25, 7, 64)      \\ \hline
TransposedConv2D & 64, 5x5, Strides 2, Pad 0, BatchNormalization, LeakyReLU  & (50, 14, 64)     \\ \hline
TransposedConv2D & 1, 5x5, Strides 2, Pad 0                                  & (100, 28, 1)     \\ \hline
\end{tabular}%
}
\caption{Generator Setting used in WGAN-GP}
\label{tab:generator}
\end{table}
\begin{table}[]
\resizebox{\linewidth}{!}{%
\begin{tabular}{|c|c|c|}
\hline
Layer   & Parameters                                         & Output Data Size \\ \hline
Input   & (100, 28, 1)                                       &                  \\ \hline
Conv2D  & 64, 5x5, Strides 2, Pad 0, LeakyReLU, Dropout 0.3  & (50, 14, 64)     \\ \hline
Conv2D  & 128, 5x5, Strides 2, Pad 0, LeakyReLU, Dropout 0.3 & (25, 7, 128)     \\ \hline
Flatten &                                                    & 22400            \\ \hline
Dense   & 256, ReLU                                          & 256              \\ \hline
Dense   & 64, ReLU                                           & 64               \\ \hline
Dense   & 1, Sigmoid                                         & 1                \\ \hline
\end{tabular}%
}
\caption{Discriminator Setting used in WGAN-GP}
\label{tab:Critic}
\end{table}
In experiments, we calculate SSIM between two spectrograms as the following:
\begin{dmath}
\label{ssim}
SSIM(x, y)=\frac{(2\mu_{x}\mu_{y}+c_1)(2\sigma_{xy}+c_2)}{(\mu_{x}^{2}+\mu_{y}^{2}+c_1)(\sigma_{x}^{2}+\sigma_{y}^{2}+c_2)}
\end{dmath}
where $x$ and $y$ are two spectrograms which need to compare structural similarity. $\mu_{x}$, $\sigma_{x}$ are the mean and variance of $x$, respectively. $\mu_{y}$, $\sigma_{x}$ are the mean and variance of $y$, respectively. And $\sigma_{xy}$ is the covariance of $x$ and $y$.  $c_1$ and $c_2$ are small constants which stabilize the division and more details can be found in~\cite{wang2003multiscale}. The range of SSIM is $[0, 1]$, where the value is closer to 1, more similar structural information they have.
\subsection{Data Collection}
To simulate a realistic scenario, we set up the experimental scene in the living room of an apartment to detect participant's activities. A PWR system and a Kinect v2 sensor~\cite{zhang2012microsoft} are synchronized and monitor the area of interest at the same time. For the PWR system, we capture the signals via two Yagi antennas- one acting as the reference antenna and another one as a surveillance antenna. The reference antenna directly captures the signal from the WiFi AP, and the surveillance antenna receives signal returns from the dynamic human subjects. Two NI USRP-2921 software defined radios are connected to the two antennas for real-time signal acquisition. The Kinect v2 sensor is used to capture the three-dimensional Mocap data of the 25 joints on the human body. Then SimHumalator can generate clean MDS according to the Mocap data. See more details in \cite{li2020passive} and \cite{vishwakarmasimhumalator}.

Three participants were used in experiments, and they completed six different tasks, including sit-down, stand-up, sit-to-walk, walk-to-sit, walk to fall-down and stand-from-floor to walk. These activities are some important in ambient assisted living and wider healthcare applications. Each activity was completed in 5--10 seconds and repeated 20 times, resulting in 60 measurements for each activity. Finally, the activity detection dataset has 360 measurement data. Correspondingly, SimHumalator generated 360 simulated datasets.

\begin{figure}
\centering
\includegraphics[width=\linewidth]{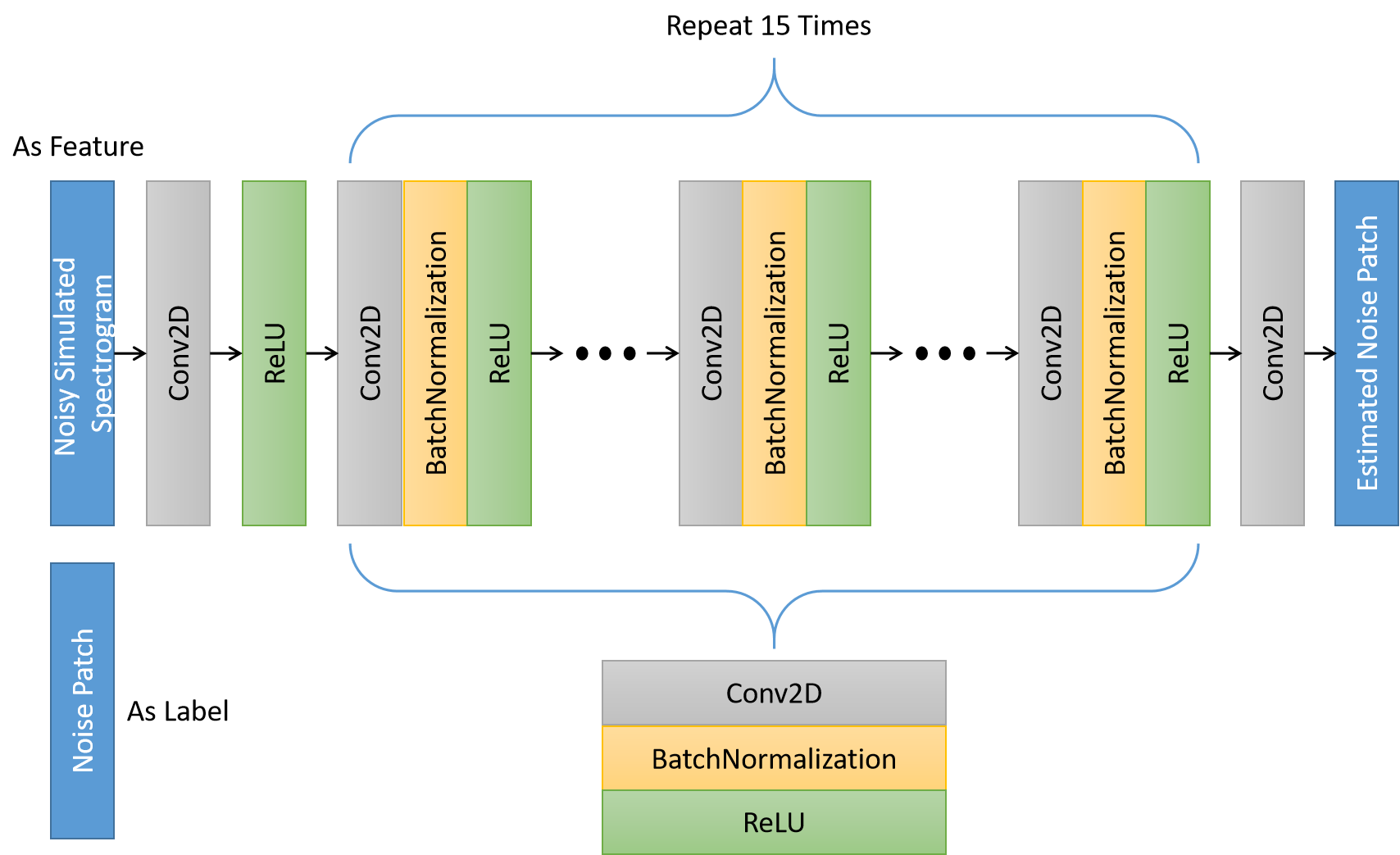}
\caption{The network details of DnCNN}
\label{fig:fig4}
\end{figure}
\subsection{Parameters Setting of Neural Networks}
There are two different neural networks in our framework. For WGAN-GP, the network overview has been shown in Fig. \ref{fig:fig3}, and the structural setting of generator and discriminator for our experiments have been shown in Table \ref{tab:generator} and Table \ref{tab:Critic}. As recommended in \cite{gulrajani2017improved}, we used RMSprop optimizer with learning rate $0.00005$ to train the GAN. For DnCNN, the structural details are shown in Fig. \ref{fig:fig4}, and it is trained with Adam optimizer with learning rate $0.01$ for $150$ epochs.
\subsection{Noise Models}
We compared the denoising performance based on three different noise models in experiments. They are:
\begin{itemize}
    \item \textbf{Measurement Noise Model (MNM)} This is our GAN-based noise model. It is learned from the real measured noise, so the noise distribution is close to the realistic scenarios. On the other hand, we can control the variance of noise patch via multiplying by the coefficient $\alpha$ i.e. $\alpha P$. In experiments, we first calculate the variance ($\sigma1$) of the clean simulated MDS and the variance ($\sigma2$) of the generated noise patch. Then according to the given noise level ($SNR$) and equation \ref{SNR}, we can obtain the desired variance ($\sigma2^{'}$) of the noise. After that, $\alpha=\frac{\sigma2^{'}}{\sigma2}$.
    \begin{dmath}
    \label{SNR}
    SNR=10*log10(\frac{\sigma_{signals}}{\sigma_{noise}})
    \end{dmath}
    \item \textbf{AWGN Model} This noise model is a common assumption in other denoising studies. In experiments, we applied zero-mean normal distribution to build this model. For controlling of the noise level, we can obtain desired noise variance according to the given noise level and equation \ref{SNR} as well. Then we can directly set the variance of the normal distribution with the calculated variance.
    \item \textbf{Mixed Noise Model (MIXM)} Based on the zero-mean AWGN model, we consider a more complex noise model, which is weighted average sampled from the AWGN model ($50\%$) and uniform distribution ($50\%$). Similarly, we can use the same method as in the AWGN noise model to control the noise level.
\end{itemize}

\section{Experiment and Results}
\label{Results Discussion}
Taking into account environmental changes, we evaluated the proposed method under eight noise levels, and tested the denoising performance on both simulated data and measured data. Meanwhile, except for DnCNN, we also trained CAE and UNet networks for comparison. The training process of UNet and CAE are different from the DnCNN architecture. They directly predict the denoised spectrograms from the given noisy input. So the label in training is changed to the clean ground truth. Theoretically, this training method may increase the amount of prediction information, thereby slowing down the model's convergence speed and increasing the cost of the training time. In experiments, we applied the similar structures introduced in \cite{de2020deep} and \cite{ronneberger2015u}, but adjusted their trainable parameters to an similar level, and limited the training time to about five minutes. In this section, we introduce these experiments and evaluate them from results qualitatively and quantitatively.
\begin{figure}
\centering
\includegraphics[width=\linewidth]{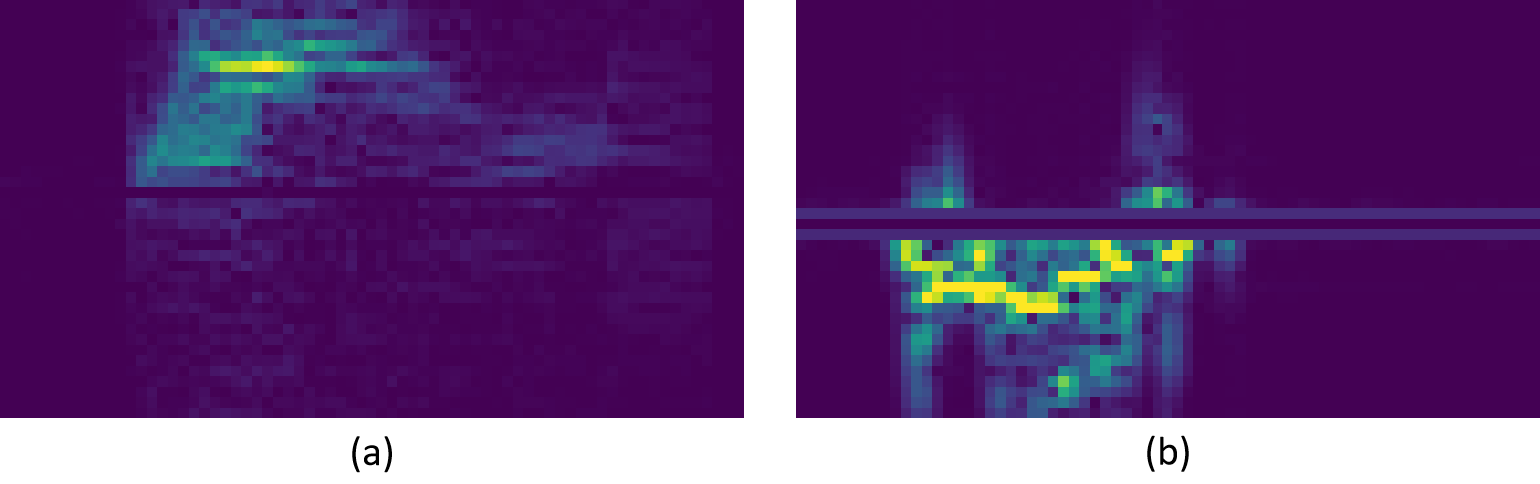}
\caption{(a): the original measured spectrogram; (b): the clean simulated spectrogram;}
\label{fig:ground truth}
\end{figure}

\subsection{Denoise Simulated MDS}
In Section \ref{Algorithm Framework} B, we create a training set for DnCNN from the clean simulated spectrograms and noise patches generated by the measurement noise model. Similarly, we can replace the MNM with AWGN or MIXM to create training sets for other two cases. After that, we derived three denoisers for the MNM, AWGN and MIXM cases. On the other hand, to test denoisers, the testing sets are created by adding clean simulated spectrograms with new noise patches. 

\subsubsection{Qualitative Analysis}
\begin{figure}
\centering
\includegraphics[width=\linewidth]{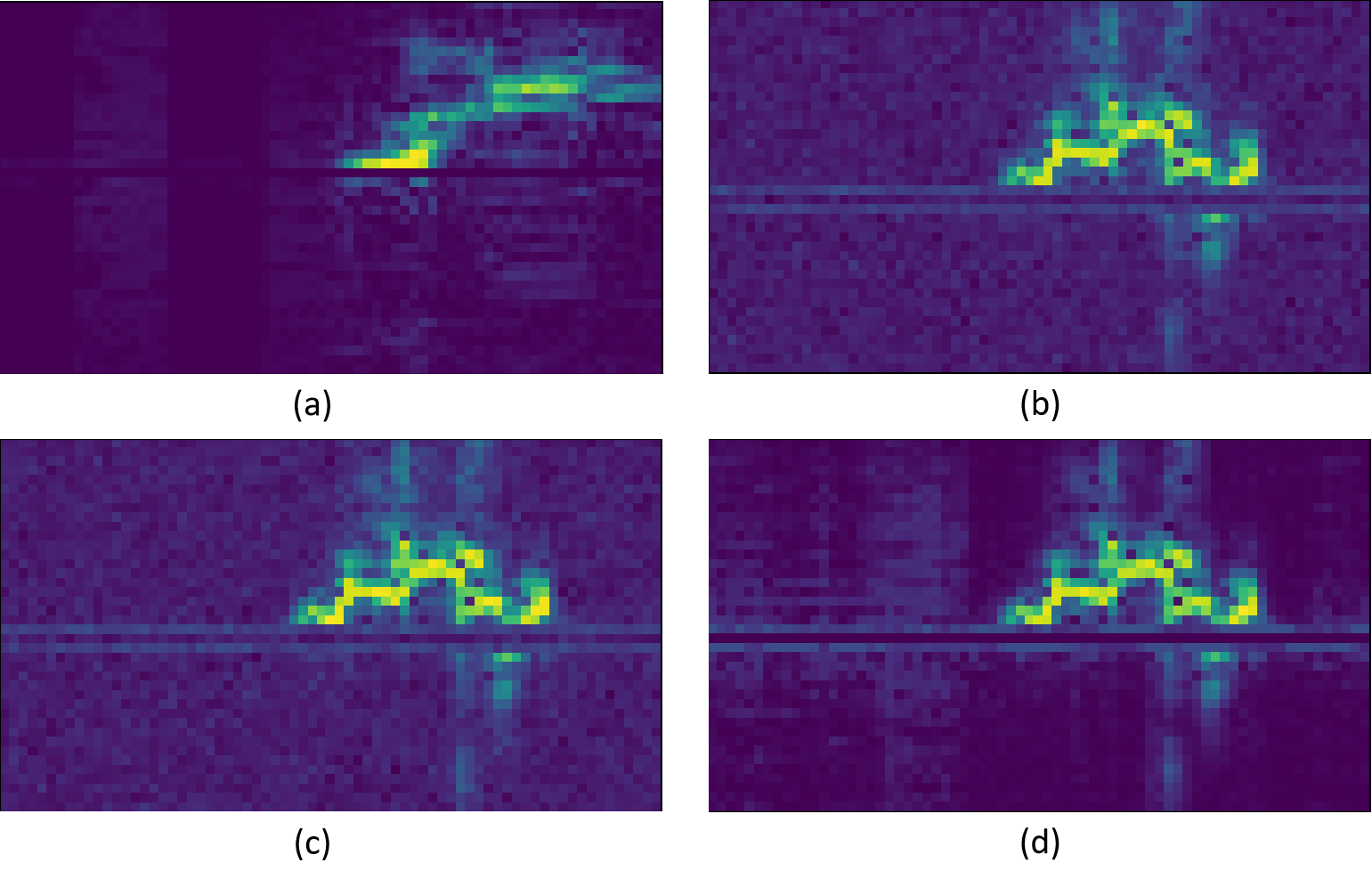}
\caption{Noisy simulated spectrograms with different noise patches: (a) is an example of the measured spectrogram; (b) simulated spectrogram plus an AWGN noise patch; (c) simulated spectrogram plus a MIXM noise patch and (d) simulated spectrogram plus a MNM noise patch}
\label{fig:noise patch}
\end{figure}

Before analyzing the denoising performance, we qualitatively evaluate which noise models can better simulate the measured noise distribution in Fig. \ref{fig:noise patch}. The plot (a) is a measured spectrogram. We observe that the measured noise is not randomly distributed across the entire spectrogram but is a narrow-rectangular pattern superimposed on some region. However, we cannot observe such patterns in plot (b) and (c) where we respectively added the AWGN noise patch and the mixed noise patch on a simulated spectrogram. This is because the AWGN noise model and mixed noise model are pixel-independent; in other words, they do not consider the spatial correlation when they generate a noise patch. By contrast, the noise distribution in the plot (d) is closer to the noise distribution of the measured spectrogram, and its noise patch is generated from the measurement noise model. Therefore, these observations indicate that our GAN-based noise modelling method can synthesize the spatial structure similar to the measured noise. For this point, our noise model is a better assumption than our noise models. This advantage can also promote the denoising performance. 

\begin{figure}
\centering
\includegraphics[width=\linewidth]{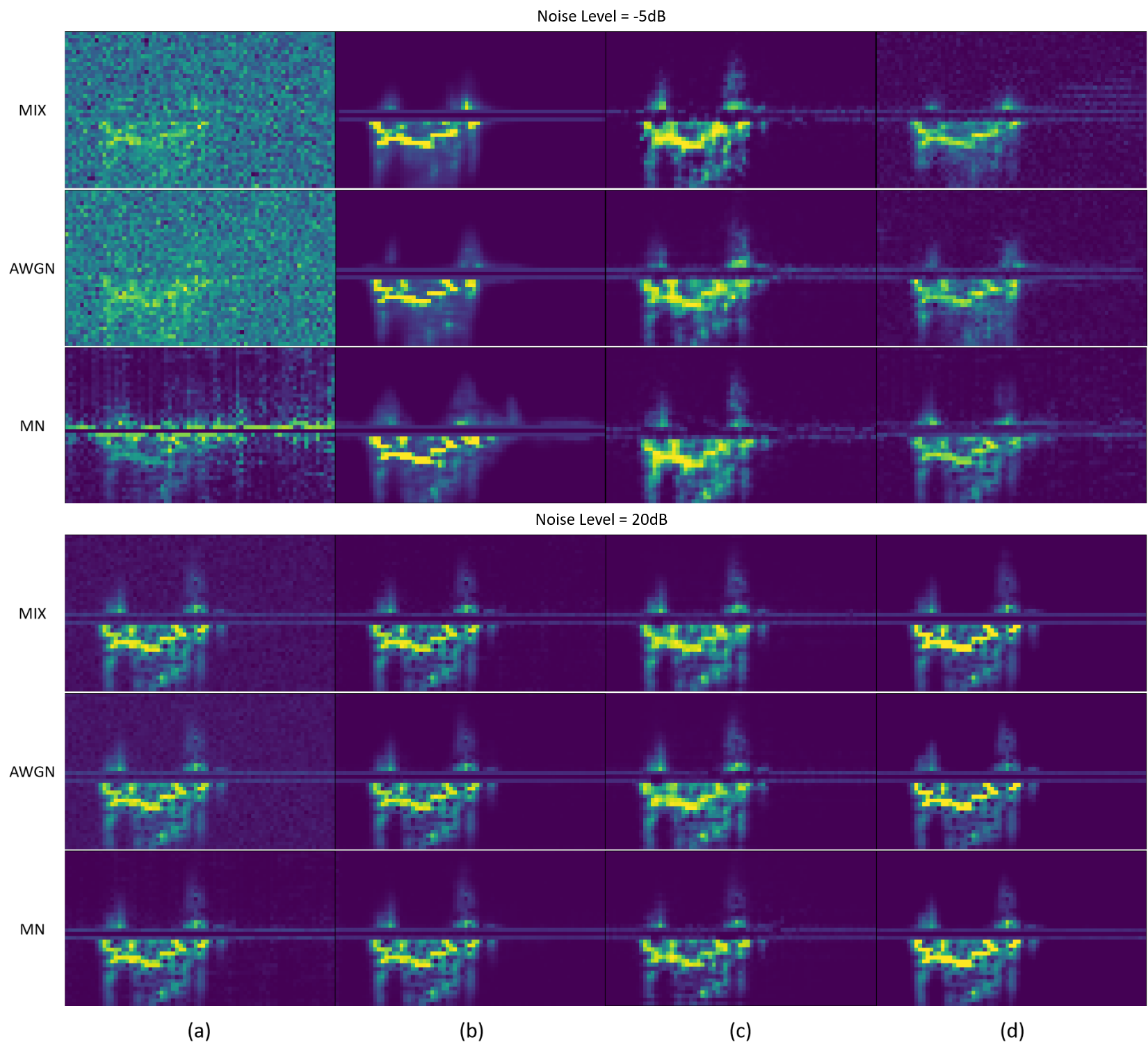}
\caption{The denoised results. column (a): the noisy simulated spectrograms; column (b): the denoised spectrograms from UNet; column (c): the denoised spectrograms from CAE; column (d): the denoised spectrograms from DnCNN; for each rows, they are different noise models cases; two noise level cases are also considered.}
\label{fig:sim qualitative result}
\end{figure}

Some denoising examples are presented in Fig. \ref{fig:sim qualitative result}, and the ground-truth spectrogram can be found in Fig. \ref{fig:ground truth} (b). Under two noise levels ($-5dB$, $20dB$), the noisy spectrogram contaminated by MN, AWGN or MIX noises are shown in column (a). Meanwhile, columns (b), (c) and (d) are outputs of denoisers trained from UNet, CAE and DnCNN, respectively. Under the $-5db$ noise level, spectrograms in column (a) are seriously contaminated. We can observe that all denoisers derived from three networks can remove most of the noises. Although, under such a low SNR condition, they lose some useful information or have distortion, denoisers derived from MNM can recover activity pattern better than two cases. For the UNet, the most significant signals can be observed, and other parts are blurred. But the MNM one has the most precise detail presentation and can recover the lower half part better than MIXM and AWGN cases. For CAE cases, although the over structure of the activity pattern is similar to the ground-truth spectrogram, the detailed information is distorted in MIX and AWGN cases. On the contrary, the MNM one is closer to the ground-truth spectrogram. For the DnCNN, it has similar results to the UNet but with the noisier background. Even though the MNM-based denoiser has better performance than MIXM and AWGN-based denoisers, we can observe those zero-value pixels in the MNM case, which are matched with the clean ground truth. Similarly, for the $20dB$ SNR condition, we can observe slight noises in MIX and AWGN cases. By contrast, the outputs in MNM do not have this issue and can obtain the cleaner denoised result. Therefore, the spatial correlation properties in our MNM can be easier learned by the neural network, which is beneficial to denoising. 

In addition, by comparing results of UNet, CAE and DnCNN, the denoised performance of the CAE denoiser is relatively worse than other denoisers. It showed fewer details than the ground truth. For the UNet and DnCNN cases, they recover the activity pattern better, and the denoised results are closer to the clean spectrogram than the CAE case. Furthermore, by comparing the signal strength i.e. the color of pixels, the pattern of DnCNN cases under the $20dB$ noise level are brighter than others. This could indicate that the DnCNN can also recover the Doppler strength better than other denoisers. We will further verify this point in quantitative analysis, and more SNR conditions will be considered.

\subsubsection{Quantitative Analysis}

\begin{table}[]

\centering
\resizebox{\linewidth}{!}{%
\begin{tabular}{@{}ccccccccccc@{}}
\cmidrule(l){3-11}
\multicolumn{2}{c}{} & \multicolumn{3}{c}{CAE} & \multicolumn{3}{c}{UNet} & \multicolumn{3}{c}{DnCNN} \\ \cmidrule(l){3-11} 
\multicolumn{2}{c}{Noise Level (dB)} & MIX & AWGN & \multicolumn{1}{c|}{MN} & MIX & AWGN & \multicolumn{1}{c|}{MN} & MIX & AWGN & MN \\ \midrule
\multirow{2}{*}{-10} & \multicolumn{1}{c|}{PSNR} & 9.43 & 10.99 & \multicolumn{1}{c|}{\textbf{19.73}} & 16.15 & 16.21 & \multicolumn{1}{c|}{19.57} & 13.85 & 14.63 & 17.94 \\
 & \multicolumn{1}{c|}{SSIM} & 0.15 & 0.32 & \multicolumn{1}{c|}{\textbf{0.74}} & 0.70 & 0.70 & \multicolumn{1}{c|}{0.73} & 0.18 & 0.19 & 0.28 \\ \midrule
\multirow{2}{*}{-5} & \multicolumn{1}{c|}{PSNR} & 6.69 & 8.29 & \multicolumn{1}{c|}{15.56} & 13.34 & 14.31 & \multicolumn{1}{c|}{\textbf{16.20}} & 12.83 & 14.08 & 16.06 \\
 & \multicolumn{1}{c|}{SSIM} & 0.32 & 0.35 & \multicolumn{1}{c|}{0.70} & 0.67 & 0.70 & \multicolumn{1}{c|}{\textbf{0.71}} & 0.32 & 0.40 & 0.43 \\ \midrule
\multirow{2}{*}{0} & \multicolumn{1}{c|}{PSNR} & 5.46 & 5.66 & \multicolumn{1}{c|}{14.28} & 11.40 & 11.05 & \multicolumn{1}{c|}{14.33} & 10.85 & 12.16 & \textbf{14.82} \\
 & \multicolumn{1}{c|}{SSIM} & 0.37 & 0.39 & \multicolumn{1}{c|}{\textbf{0.69}} & 0.64 & 0.64 & \multicolumn{1}{c|}{0.67} & 0.36 & 0.43 & 0.58 \\ \midrule
\multirow{2}{*}{5} & \multicolumn{1}{c|}{PSNR} & 3.66 & 6.16 & \multicolumn{1}{c|}{9.94} & 6.14 & 9.85 & \multicolumn{1}{c|}{9.99} & 9.40 & 10.04 & \textbf{12.5} \\
 & \multicolumn{1}{c|}{SSIM} & 0.39 & 0.50 & \multicolumn{1}{c|}{\textbf{0.59}} & 0.52 & 0.54 & \multicolumn{1}{c|}{0.58} & 0.32 & 0.39 & 0.51 \\ \midrule
\multirow{2}{*}{10} & \multicolumn{1}{c|}{PSNR} & 0.00 & 4.32 & \multicolumn{1}{c|}{5.36} & 3.55 & 4.43 & \multicolumn{1}{c|}{5.84} & 7.97 & 7.69 & \textbf{9.85} \\
 & \multicolumn{1}{c|}{SSIM} & 0.35 & 0.46 & \multicolumn{1}{c|}{\textbf{0.47}} & 0.39 & 0.42 & \multicolumn{1}{c|}{\textbf{0.47}} & 0.28 & 0.38 & 0.44 \\ \midrule
\multirow{2}{*}{15} & \multicolumn{1}{c|}{PSNR} & -2.99 & -2.32 & \multicolumn{1}{c|}{0.0} & 1.82 & 3.59 & \multicolumn{1}{c|}{3.92} & 6.00 & 8.14 & \textbf{11.61} \\
 & \multicolumn{1}{c|}{SSIM} & 0.20 & 0.30 & \multicolumn{1}{c|}{0.31} & 0.22 & 0.28 & \multicolumn{1}{c|}{0.32} & 0.21 & 0.32 & \textbf{0.34} \\ \midrule
\multirow{2}{*}{20} & \multicolumn{1}{c|}{PSNR} & -7.07 & -6.12 & \multicolumn{1}{c|}{-5.64} & -4.05 & -1.2 & \multicolumn{1}{c|}{0.87} & 5.88 & 6.43 & \textbf{8.40} \\
 & \multicolumn{1}{c|}{SSIM} & 0.07 & 0.15 & \multicolumn{1}{c|}{0.19} & 0.07 & 0.14 & \multicolumn{1}{c|}{0.19} & 0.12 & 0.14 & \textbf{0.21} \\ \midrule
\multirow{2}{*}{25} & \multicolumn{1}{c|}{PSNR} & -13.46 & -12.71 & \multicolumn{1}{c|}{-12.57} & -7.23 & -5.2 & \multicolumn{1}{c|}{-3.19} & 4.21 & 6.19 & \textbf{6.71} \\
 & \multicolumn{1}{c|}{SSIM} & 0.00 & 0.00 & \multicolumn{1}{c|}{0.04} & 0.0 & 0.04 & \multicolumn{1}{c|}{0.07} & 0.03 & 0.09 & \textbf{0.12} \\ \bottomrule
\end{tabular}%
}
\caption{\label{tab: sim quantitative results}The improvements of PSNR and SSIM for different activities of CAE, UNet and DnCNN networks. It also includes MIXM, AWGN and MNM cases and eight different noise levels.}
\end{table}

\begin{table}[]

\resizebox{\linewidth}{!}{%
\begin{tabular}{cccccccc}
\hline
\multicolumn{2}{c}{Activities}                    & sit-down & stand-up & sit-to-walk & walk-to-sit & walk-to-fall & stand-from-floor \\ \hline
Noise Level           & \multicolumn{7}{c}{SNR=-5dB}                                              \\ \hline
\multirow{2}{*}{MIX}  & \multicolumn{1}{c|}{PSNR} & 12.79 & 13.05 & 12.92 & 12.96 & 12.65 & 12.62 \\
                      & \multicolumn{1}{c|}{SSIM} & 0.32  & 0.32  & 0.33  & 0.33  & 0.32  & 0.33  \\
\multirow{2}{*}{AWGN} & \multicolumn{1}{c|}{PSNR} & 14.21    & 14.36    & 14.16       & 14.29       & 13.74        & 13.82            \\
                      & \multicolumn{1}{c|}{SSIM} & 0.41  & 0.41  & 0.40  & 0.41  & 0.40  & 0.40  \\
\multirow{2}{*}{MN}   & \multicolumn{1}{c|}{PSNR} & 16.24 & 16.25 & 16.25 & 16.09 & 15.86 & 15.65 \\
                      & \multicolumn{1}{c|}{SSIM} & 0.41  & 0.44  & 0.43  & 0.43  & 0.44  & 0.41  \\ \hline
Noise Level           & \multicolumn{7}{c}{SNR=10dB}                                              \\ \hline
\multirow{2}{*}{MIX}  & \multicolumn{1}{c|}{PSNR} & 8.06  & 8.15  & 8.08  & 8.21  & 7.71  & 7.62  \\
                      & \multicolumn{1}{c|}{SSIM} & 0.30  & 0.30  & 0.29  & 0.26  & 0.27  & 0.27  \\
\multirow{2}{*}{AWGN} & \multicolumn{1}{c|}{PSNR} & 7.79  & 7.87  & 7.84  & 7.96  & 7.40  & 7.28  \\
                      & \multicolumn{1}{c|}{SSIM} & 0.41  & 0.41  & 0.40  & 0.39  & 0.35  & 0.34  \\
\multirow{2}{*}{MN}   & \multicolumn{1}{c|}{PSNR} & 10.20 & 10.23 & 9.99  & 9.70  & 9.57  & 9.37  \\
                      & \multicolumn{1}{c|}{SSIM} & 0.47  & 0.47  & 0.45  & 0.44  & 0.41  & 0.39  \\ \hline
Noise Level           & \multicolumn{7}{c}{SNR=20dB}                                              \\ \hline
\multirow{2}{*}{MIX}  & \multicolumn{1}{c|}{PSNR} & 6.51  & 6.49  & 6.51  & 6.59  & 5.99  & 5.81  \\
                      & \multicolumn{1}{c|}{SSIM} & 0.12  & 0.12  & 0.11  & 0.10  & 0.11  & 0.11  \\
\multirow{2}{*}{AWGN} & \multicolumn{1}{c|}{PSNR} & 5.67  & 5.46  & 5.70  & 6.77  & 5.87  & 5.82  \\
                      & \multicolumn{1}{c|}{SSIM} & 0.18  & 0.17  & 0.18  & 0.12  & 0.12  & 0.11  \\
\multirow{2}{*}{MN}   & \multicolumn{1}{c|}{PSNR} & 8.71  & 8.53  & 8.37  & 8.23  & 8.14  & 7.95  \\
                      & \multicolumn{1}{c|}{SSIM} & 0.25  & 0.25  & 0.24  & 0.23  & 0.21  & 0.20  \\ \hline
\end{tabular}%
}
\caption{\label{tab: sim different activities results}The improvements of PSNR and SSIM for different activities. The results are based on DnCNN and tested under three SNR levels.}
\end{table}
In the quantitative analysis, SSIM is the main focus used to evaluate the denoising performance, and PSNR is also calculated in the result. Specifically, we considered the improvements of these indexes before and after denoising. For the given clean simulated ground truth $gt$, the noisy simulated spectrogram $noisyS$ and the denoised spectrogram $denoisedS$, we first calculate the PSNR and SSIM between $gt$ and $noisyS$ i.e. $PSNR_{noisy}(gt, noisyS)$ and $SSIM_{noisy}(gt, noisyS)$. Then we calculated $PSNR_{denoised}(gt, denoisedS)$ and $SSIM_{denoised}(gt, denoisedS)$. At the end, the difference between them are the improvements of PSNR and SSIM. We used this process to calculate quantitative results for all cases and summarized results in Table \ref{tab: sim quantitative results}. 

The table shows the results of CAE, UNet and DnCNN. Meanwhile, different denoisers obtained from MNM, AWGN and MIXM are also included for each networks. Furthermore, for each noise levels, we bolded the maximum improvements. Then, we can observe the following phenomenons:
\begin{enumerate}
    \item The MNM-based denoising performance is always better for any networks than others while the MIXM-based case is the worst one. This is because that the AWGN and MIX are pixel-independent noises, which is not realistic and also challenging for a neural network to learn the correlation. Especially, MIXM is more complicated than AWGN, so the case has worse results. On the contrary, our MNM considers the spatial correlation of noises and learned them from the real measured MDS, which helps train a CNN denoiser.
    \item For most of cases, with the increase of the SNR, the improvements of PSNR and SSIM decrease, but generally, they are positive changes. However, in some cases under CAE and UNet, the changes of PSNR become negative and the improvements of SSIM dropped to $0$. This shows that with the improvement of SNR, these two networks can no longer effectively improve MDS quality, which is obviously shown in the changes of PSNR value. However, even so, MNM-based denoisers can slightly improve the SSIM, which means that the model is still optimising the structural information of the noisy spectrogram.
    \item On the other hand, DnCNN denoisers do not show any negative changes in the higher SNR conditions and still perform well. This difference may be due to their training methods. As a model that focuses on predicting noise patch, DnCNN may change the original information less. However, we can also observe that the other two networks show the better results in the low SNR conditions, e.g. $-10dB$. This shows that in a complex environment, it may be a better way to estimate useful information directly.
\end{enumerate}

Meanwhile, we separately show the denoising performance based on DnCNN for different activities under Bad($-5dB$), Mild($10dB$) and Good($20dB$) noise conditions in Table \ref{tab: sim different activities results}. For each activity column, we can observe similar phenomenons, as presented above. Moreover, compared between different activities, the denoising performance is stable all the time. This shows the robustness of our method.

According to these analysis, our MNM is a better modelling choice in terms of improving denoising performance. We will further demonstrate this point in the following analysis of the measured data. On the other hand, the experimental result also shows that we can choose a more suitable network according to the different noise levels.

\subsection{Denoise Measured MDS}
Unlike the simulated data, we are not able to obtain the clean measured spectrograms from experiments. As a result, we will reflect the denoising performance from the activity classification accuracy. An effective denoiser can remove noises while retaining important features, so that we can get higher classification results. In this section, the previous denoisers trained by simulated spectrograms are still used. But the testing set now is our noisy measured spectrograms. An example of the noisy measured spectrogram is shown in Fig. \ref{fig:ground truth} (a), where results presented in the qualitative analysis are based on this spectrogram.
\subsubsection{Qualitative Analysis}
\begin{figure}
\centering
\includegraphics[width=\linewidth]{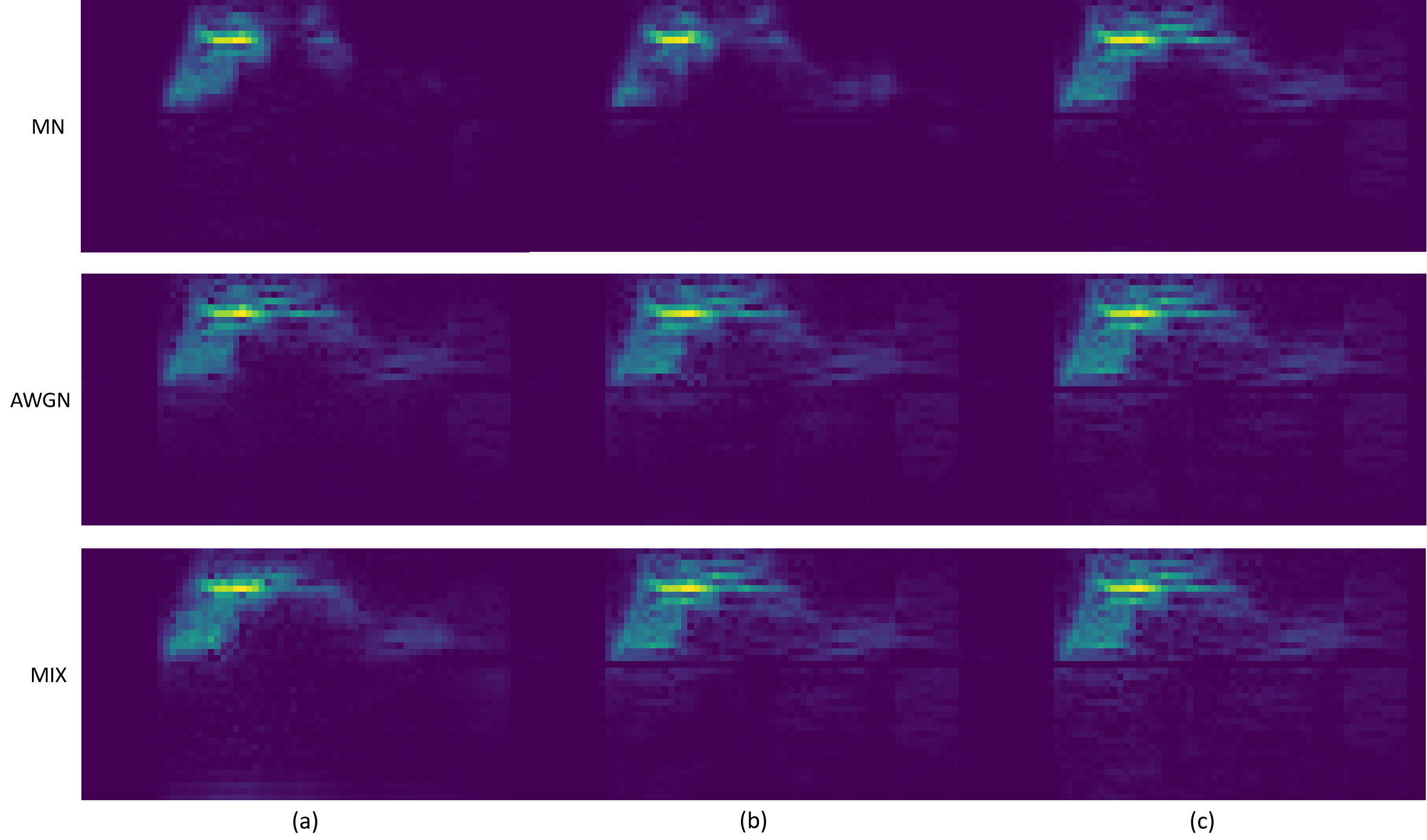}
\caption{The MNM, AWGN and MIXM-based DnCNN denoised results. column (a): $-5dB$ cases; column (b): $10dB$ cases; column (c): $20dB$ cases}
\label{fig:mear qualitative}
\end{figure}
From Fig. \ref{fig:ground truth} (a), we can observe that, in addition to the noise superimposed on the activity signal, the noise interference in the lower half and the right side is quite obvious. Removing these noises will help improve the quality of MDS. In qualitative analysis, we presented some results of the DnCNN denoisers under different noise conditions. Three noise models-based situations are considered. In Fig. \ref{fig:mear qualitative}, each rows show the denoised spectrograms obtained from the MNM, AWGN or MIXM-based denoiser, and columns (a)(b)(c) are Bad, Mild and Good noise condition cases.

Firstly, for Bad cases, the denoisers are trained with quite low SNR spectrograms. Compared the measured MDS with the previous examples, it has the much better SNR. Therefore, the denoisers might overestimate the value of the noise and causes some useful information of results in (a) is removed. From the noise perspective, the denoised spectrogram from MNM case effectively removed the noise in the lower half and on the far right. However, in the other two cases, even though the noise in the lower half is mitigated, the noise on the far right of the spectrogram is still relatively clear. This means that these two models cannot detect and weaken the noise of special shapes well. They just adjust the value for each pixel without considering spatial correlation. This weakness is more and more obvious in Mild and Good conditions for AWGN and MIXM cases. These results can not only observe the noise on the right, but also the noise in the lower half cannot be effectively removed. By contrast, although the MNM case removed more useful information at the Bad noise condition, in the case of closer to the real noise level, it not only retains clear motion information, but also effectively removes obvious noise. 

On the basis of these observations, we can initially decide that our denoising framework is feasible and can be applied to the measured data. MNM, meanwhile, is preferable to other forms of noise. To efficiently minimize noise, it can thoroughly understand the spatial correlation of noise in real situations. From the classification performance in the next part, these conclusions will be further validated.

\subsubsection{Quantitative Analysis}
As mentioned before, the clean measured data could not be collected for comparison, so the PSNR and SSIM analysis is not available. We therefore indirectly show the denoising efficiency from the accuracy of the classifying denoised spectrograms, where the classifier that we have implemented is VGG16\cite{simonyan2014very}. Under eight noise levels, we consider CAE, UNet and DnCNN-based denoizers for MNM, AWGN and MIXM cases. The classification results are summarized in Table \ref{tab: class results}, and the bold values are the highest classification results for each noise levels. 

\begin{figure}
\centering
\includegraphics[width=\linewidth]{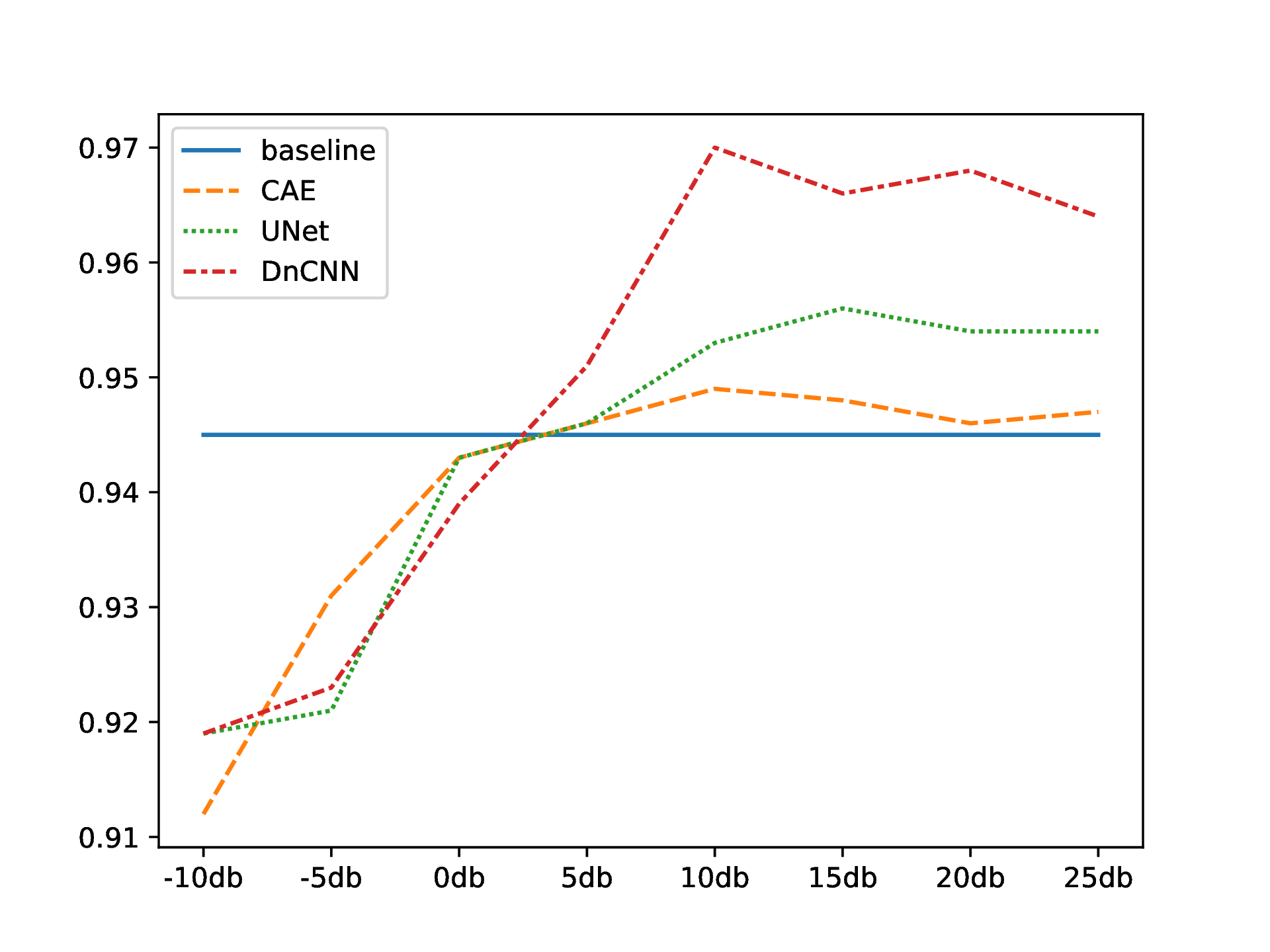}
\caption{The classification result comparison between CAE, UNet, DnCNN and the baseline}
\label{fig: MNM class results}
\end{figure}

We can observe the following phenomenons from the table:
\begin{enumerate}
    \item For each networks, the classification results obtained by the denoiser trained at various noise levels are different. In low SNR situations, denoisers may exclude some useful signals from the spectrogram that result in lower classification accuracy. However, as the level of noise increases, the accuracy is not always increased. The peak values for three networks are seen in the case of $10dB$ or $15dB$. This means that the noise level of the measured data is likely to be within this range, and it also shows that when the noise level used in training a denoiser matches the real environmental noise level, the better denoising performance can be obtained.
    \item For a clearer comparison, we plot classification results of the MNM case for three networks in Fig. \ref{fig: MNM class results}. The blue line is the baseline accuracy obtained by directly classifying the original measured spectrograms, which is around $94.5\%$. From the figure, we can observe that before the $5dB$, the denoised spectrograms cannot help improve the classification accuracy, which is because that denoisers trained under such the low SNR conditions removed useful features. Then, as SNR approaches the real situation, the accuracy is significantly increased. Especially, in the case of $10dB$, the DnCNN denoiser derived from the MNM obtains the highest classification accuracy. This finding is around 2\% higher than the performance of MNM in other networks. After that, the accuracy of all three networks dropped slightly, but was still better than the baseline value. 
    \item In addition, we also investigated the effects caused by the different number of the training set. The split rate, $\frac{m}{1-m}$, represents the ratio between the number of training spectrograms and the number of testing spectrograms, where $m$ is the proportion of the training set. From the table, the higher split rate the better classification accuracy for all cases. However, for the radar sensing tasks, we often face training data shortages that will reduce the efficiency of the system. But we can also observe that under $10dB$ to $25dB$ cases, the DnCNN denoisers derived by the MNM has better results than baseline in most of split rates. It indicates that the proposed framework can not only reduce noise well, but also effectively help improve accuracy when the training set is insufficient.
\end{enumerate}

Once again, we have confirmed that the proposed framework is efficient for real measured spectrograms, and more specifically, compared to other noise models, the MNM has superiority. Under the appropriate noise level, efficiency exceeds AWGN and MIXM and, in the event of inadequate training data, significant improvements can still be obtained.

\begin{table*}[]
\centering
\resizebox{.8\textwidth}{!}{%
\begin{tabular}{@{}cclllllllll@{}}
\cmidrule(l){3-11}
\multicolumn{2}{c}{} & \multicolumn{3}{c}{CAE} & \multicolumn{3}{c}{UNet} & \multicolumn{3}{c}{DnCNN} \\ \midrule
Noise Level (dB) & \multicolumn{1}{c|}{Split Rate} & \multicolumn{1}{c}{AWGN} & \multicolumn{1}{c}{MIX} & \multicolumn{1}{c|}{MN} & \multicolumn{1}{c}{AWGN} & \multicolumn{1}{c}{MIX} & \multicolumn{1}{c|}{MN} & \multicolumn{1}{c}{AWGN} & \multicolumn{1}{c}{MIX} & \multicolumn{1}{c}{MN} \\ \midrule
\multirow{3}{*}{-10} & \multicolumn{1}{c|}{0.8/0.2} & \multicolumn{1}{c}{0.908} & \multicolumn{1}{c}{0.894} & \multicolumn{1}{c|}{0.912} & \multicolumn{1}{c}{0.916} & \multicolumn{1}{c}{0.902} & \multicolumn{1}{c|}{\textbf{0.919}} & \multicolumn{1}{c}{0.906} & \multicolumn{1}{c}{0.889} & \multicolumn{1}{c}{\textbf{0.919}} \\
 & \multicolumn{1}{c|}{0.7/0.3} & \multicolumn{1}{c}{0.900} & \multicolumn{1}{c}{0.850} & \multicolumn{1}{c|}{0.910} & \multicolumn{1}{c}{0.903} & \multicolumn{1}{c}{0.902} & \multicolumn{1}{c|}{0.911} & \multicolumn{1}{c}{0.904} & \multicolumn{1}{c}{0.973} & \multicolumn{1}{c}{0.906} \\
 & \multicolumn{1}{c|}{0.6/0.4} & \multicolumn{1}{c}{0.893} & \multicolumn{1}{c}{0.841} & \multicolumn{1}{l|}{0.897} & \multicolumn{1}{c}{0.903} & \multicolumn{1}{c}{0.897} & \multicolumn{1}{l|}{0.903} & \multicolumn{1}{c}{0.896} & \multicolumn{1}{c}{0.823} & \multicolumn{1}{c}{0.895} \\ \midrule
\multirow{3}{*}{-5} & \multicolumn{1}{c|}{0.8/0.2} & \multicolumn{1}{c}{0.903} & \multicolumn{1}{c}{0.904} & \multicolumn{1}{c|}{\textbf{0.931}} & \multicolumn{1}{c}{0.918} & \multicolumn{1}{c}{0.908} & \multicolumn{1}{l|}{0.921} & \multicolumn{1}{c}{0.916} & \multicolumn{1}{c}{0.902} & \multicolumn{1}{c}{0.923} \\
 & \multicolumn{1}{c|}{0.7/0.3} & \multicolumn{1}{c}{0.900} & \multicolumn{1}{c}{0.850} & \multicolumn{1}{l|}{0.910} & \multicolumn{1}{c}{0.903} & \multicolumn{1}{c}{0.902} & \multicolumn{1}{l|}{0.911} & \multicolumn{1}{c}{0.904} & \multicolumn{1}{c}{0.873} & \multicolumn{1}{c}{0.906} \\
 & \multicolumn{1}{c|}{0.6/0.4} & \multicolumn{1}{c}{0.893} & \multicolumn{1}{c}{0.941} & \multicolumn{1}{l|}{0.897} & \multicolumn{1}{c}{0.903} & \multicolumn{1}{c}{0.897} & \multicolumn{1}{l|}{0.903} & \multicolumn{1}{c}{0.896} & \multicolumn{1}{c}{0.923} & \multicolumn{1}{c}{0.895} \\ \midrule
\multirow{3}{*}{0} & \multicolumn{1}{c|}{0.8/0.2} & \multicolumn{1}{c}{0.920} & \multicolumn{1}{c}{0.917} & \multicolumn{1}{l|}{\textbf{0.943}} & \multicolumn{1}{c}{0.936} & \multicolumn{1}{c}{0.925} & \multicolumn{1}{l|}{\textbf{0.943}} & \multicolumn{1}{c}{0.932} & \multicolumn{1}{c}{0.917} & \multicolumn{1}{c}{0.939} \\
 & \multicolumn{1}{c|}{0.7/0.3} & \multicolumn{1}{c}{0.901} & \multicolumn{1}{c}{0.899} & \multicolumn{1}{l|}{0.920} & \multicolumn{1}{c}{0.930} & \multicolumn{1}{c}{0.912} & \multicolumn{1}{l|}{0.938} & \multicolumn{1}{c}{0.930} & \multicolumn{1}{c}{0.907} & \multicolumn{1}{c}{0.935} \\
 & \multicolumn{1}{c|}{0.6/0.4} & \multicolumn{1}{c}{0.897} & \multicolumn{1}{c}{0.885} & \multicolumn{1}{l|}{0.909} & \multicolumn{1}{c}{0.916} & \multicolumn{1}{c}{0.908} & \multicolumn{1}{l|}{0.926} & \multicolumn{1}{c}{0.913} & \multicolumn{1}{c}{0.922} & \multicolumn{1}{c}{0.928} \\ \midrule
\multirow{3}{*}{5} & \multicolumn{1}{c|}{0.8/0.2} & \multicolumn{1}{c}{0.938} & \multicolumn{1}{c}{0.919} & \multicolumn{1}{l|}{0.946} & \multicolumn{1}{c}{0.940} & \multicolumn{1}{c}{0.919} & \multicolumn{1}{l|}{0.946} & \multicolumn{1}{c}{0.942} & \multicolumn{1}{c}{0.933} & \textbf{0.951} \\
 & \multicolumn{1}{c|}{0.7/0.3} & \multicolumn{1}{c}{0.911} & \multicolumn{1}{c}{0.903} & \multicolumn{1}{l|}{0.921} & \multicolumn{1}{c}{0.933} & \multicolumn{1}{c}{0.925} & \multicolumn{1}{l|}{0.935} & \multicolumn{1}{c}{0.931} & \multicolumn{1}{c}{0.928} & \multicolumn{1}{c}{0.944} \\
 & \multicolumn{1}{c|}{0.6/0.4} & \multicolumn{1}{c}{0.896} & \multicolumn{1}{c}{0.897} & \multicolumn{1}{l|}{0.911} & \multicolumn{1}{c}{0.914} & \multicolumn{1}{c}{0.919} & \multicolumn{1}{l|}{0.925} & \multicolumn{1}{c}{0.929} & \multicolumn{1}{c}{0.911} & \multicolumn{1}{c}{0.934} \\ \midrule
\multirow{3}{*}{10} & \multicolumn{1}{c|}{0.8/0.2} & \multicolumn{1}{c}{0.936} & \multicolumn{1}{c}{0.928} & \multicolumn{1}{l|}{0.949} & \multicolumn{1}{c}{0.949} & \multicolumn{1}{c}{0.939} & \multicolumn{1}{l|}{0.953} & \multicolumn{1}{c}{0.952} & \multicolumn{1}{c}{0.948} & \textbf{0.970} \\
 & \multicolumn{1}{c|}{0.7/0.3} & \multicolumn{1}{c}{0.919} & \multicolumn{1}{c}{0.912} & \multicolumn{1}{l|}{0.923} & \multicolumn{1}{c}{0.932} & \multicolumn{1}{c}{0.932} & \multicolumn{1}{l|}{0.946} & \multicolumn{1}{c}{0.946} & \multicolumn{1}{c}{0.948} & \multicolumn{1}{c}{0.963} \\
 & \multicolumn{1}{c|}{0.6/0.4} & \multicolumn{1}{c}{0.903} & \multicolumn{1}{c}{0.903} & \multicolumn{1}{l|}{0.913} & \multicolumn{1}{c}{0.926} & \multicolumn{1}{c}{0.928} & \multicolumn{1}{l|}{0.936} & \multicolumn{1}{c}{0.938} & \multicolumn{1}{c}{0.933} & \multicolumn{1}{c}{0.944} \\ \midrule
\multirow{3}{*}{15} & \multicolumn{1}{c|}{0.8/0.2} & \multicolumn{1}{c}{0.945} & \multicolumn{1}{c}{0.926} & \multicolumn{1}{l|}{0.948} & \multicolumn{1}{c}{0.942} & \multicolumn{1}{c}{0.945} & \multicolumn{1}{l|}{0.956} & \multicolumn{1}{c}{0.950} & \multicolumn{1}{c}{0.949} & \textbf{0.966} \\
 & \multicolumn{1}{c|}{0.7/0.3} & \multicolumn{1}{c}{0.923} & \multicolumn{1}{c}{0.920} & \multicolumn{1}{l|}{0.928} & \multicolumn{1}{c}{0.935} & \multicolumn{1}{c}{0.942} & \multicolumn{1}{l|}{0.950} & \multicolumn{1}{c}{0.942} & \multicolumn{1}{c}{0.941} & \textbf{0.966} \\
 & \multicolumn{1}{c|}{0.6/0.4} & \multicolumn{1}{c}{0.917} & \multicolumn{1}{c}{0.909} & \multicolumn{1}{l|}{0.920} & \multicolumn{1}{c}{0.930} & \multicolumn{1}{c}{0.933} & \multicolumn{1}{l|}{0.944} & \multicolumn{1}{c}{0.939} & \multicolumn{1}{c}{0.930} & \multicolumn{1}{c}{0.951} \\ \midrule
\multirow{3}{*}{20} & \multicolumn{1}{c|}{0.8/0.2} & \multicolumn{1}{c}{0.932} & \multicolumn{1}{c}{0.925} & \multicolumn{1}{l|}{0.946} & \multicolumn{1}{c}{0.945} & \multicolumn{1}{c}{0.941} & \multicolumn{1}{l|}{0.954} & \multicolumn{1}{c}{0.951} & \multicolumn{1}{c}{0.943} & \textbf{0.968} \\
 & \multicolumn{1}{c|}{0.7/0.3} & \multicolumn{1}{c}{0.925} & \multicolumn{1}{c}{0.913} & \multicolumn{1}{l|}{0.934} & \multicolumn{1}{c}{0.934} & \multicolumn{1}{c}{0.939} & \multicolumn{1}{l|}{0.947} & \multicolumn{1}{c}{0.946} & \multicolumn{1}{c}{0.939} & \multicolumn{1}{c}{0.964} \\
 & \multicolumn{1}{c|}{0.6/0.4} & \multicolumn{1}{c}{0.921} & \multicolumn{1}{c}{0.896} & \multicolumn{1}{l|}{0.930} & \multicolumn{1}{c}{0.927} & \multicolumn{1}{c}{0.931} & \multicolumn{1}{l|}{0.941} & \multicolumn{1}{c}{0.937} & \multicolumn{1}{c}{0.931} & \multicolumn{1}{c}{0.948} \\ \midrule
\multirow{3}{*}{25} & \multicolumn{1}{c|}{0.8/0.2} & \multicolumn{1}{c}{0.935} & \multicolumn{1}{c}{0.925} & \multicolumn{1}{l|}{0.947} & \multicolumn{1}{c}{0.941} & \multicolumn{1}{c}{0.942} & \multicolumn{1}{l|}{0.954} & \multicolumn{1}{c}{0.945} & \multicolumn{1}{c}{0.942} & \textbf{0.964} \\
 & \multicolumn{1}{c|}{0.7/0.3} & \multicolumn{1}{c}{0.925} & \multicolumn{1}{c}{0.916} & \multicolumn{1}{l|}{0.935} & \multicolumn{1}{c}{0.928} & \multicolumn{1}{c}{0.937} & \multicolumn{1}{l|}{0.945} & \multicolumn{1}{c}{0.945} & \multicolumn{1}{c}{0.942} & \textbf{0.964} \\
 & \multicolumn{1}{c|}{0.6/0.4} & \multicolumn{1}{c}{0.909} & \multicolumn{1}{c}{0.918} & \multicolumn{1}{l|}{0.929} & \multicolumn{1}{c}{0.916} & \multicolumn{1}{c}{0.925} & \multicolumn{1}{l|}{0.939} & \multicolumn{1}{c}{0.934} & \multicolumn{1}{c}{0.927} & \multicolumn{1}{c}{0.945} \\ \bottomrule
\end{tabular}%
}
\caption{\label{tab: class results}The classification results}
\end{table*}

\section{Conclusion}
\label{Conclusion}
In this paper, we present a novel MDS denoising framework. Different from previous DNN-based methods which use AWGN model to create the training set, we proposed a GAN-based real noise modelling method. The obtained MNM can better synthesize the real noise distribution and improve the denoising performance. Various experiments have been done on both the simulated data and the measured data. Through qualitative and quantitative analysis, we can verify the performance of our denoising framework, and more importantly, the proposed method has been compared with networks, such as CAE and UNet, which can validated that the MNM plus DnCNN system is a better choice for Micro-Doppler spectrogram denoising. In addition, by comparing the results of classification, our approach can be applied on the basis of measured data and can perform well even in the event of a lack of training set, which is of significant importance. One question we did not discuss in this paper is: does the previous measurement noise model still work when the environment changes? We think the answer depends on how much change it has. For small change, the previous model might still be effective. However, for huge change, we need to retrain a new measurement noise model. In the future, we plan to consider this aspect and improve our framework to adapt to various environmental changes.
\section*{Acknowledgments}
This work is part of the OPERA project funded by the UK Engineering and Physical Sciences Research Council (EPSRC), Grant No: EP/R018677/1. 


\bibliographystyle{IEEEtran}

\bibliography{refs.bib}

\balance

\end{document}